# Enhanced Chest Disease Classification Using an Improved CheXNet Framework with EfficientNetV2-M and Optimization-Driven Learning

**Ali M. Bahram**[1], **Saman Muhammad Omer**[2], **Hardi M. Mohammed**[3], **Sirwan Abdolwahed Aula**[4]

[1]MSc Student at Charmo University Department of Computer Science and Computer Lab Technician at Department of Network Technology, Computer Science Institute (CSI) at Sulaimani Polytechnic University, Sulaimani, Kurdistan Region, Iraq

[2]PhD, Assistant Professor, Department of Software Engineering, University of Raparin, Raparin, as Sulaymaniyah, Kurdistan Region, Iraq

[3]PhD, Lecturer at Computer Science Department, College of Science, Charmo University, Chamchamal as Sulaymaniyah, Sulaimani, Kurdistan Region, Iraq

[4]Department of Computer Science, Faculty of Science, Soran University, Kurdistan Region, Iraq

**Corresponding Author**: Ali M. Bahram.

Email: ali.m.bahram@spu.edu.iq.

ORCID: https://orcid.org/0009-0009-5777-5134

**Supervisor**: Saman Muhammad Omer.

Email: saman.muhammad@uor.edu.krd

ORCID: https://orcid.org/0000-0002-5517-9066

**Co-supervisor**: Hardi M. Mohammed.

Email: hardi.mohammed@chu.edu.iq

ORCID: https://orcid.org/0000-0002-9766-9100

**Co-supervisor**: Sirwan Abdolwahed Aula.

Email: sirwan.aula@soran.edu.iq

ORCID: https://orcid.org/0009-0001-5665-3524

## ABSTRACT

*The interpretation of Chest X-ray is an important diagnostic issue in clinical practice and especially in the resource-limited setting where the shortage of radiologists plays a role in delayed diagnosis and poor patient outcomes. Although the original CheXNet architecture has shown potential in automated analysis of chest radiographs, DenseNet-121 backbone is computationally inefficient and poorly single-label classifier. To eliminate such shortcomings, we suggest a better classification framework of chest disease that relies on EfficientNetV2-M and incorporates superior training approaches such as Automatic Mixed Precision training, AdamW, Cosine Annealing learning rate scheduling, and Exponential Moving Average regularization. We prepared a dataset of 18,080 chest X-ray images of three source materials of high authority and representing five key clinically significant disease categories which included Cardiomegaly, COVID-19, Normal, Pneumonia, and Tuberculosis. To achieve statistical reliability and reproducibility, nine independent experimental runs were run. The suggested architecture showed significant gains and the mean test accuracy of 96.45% ± 0.17% compared to 95.30% ± 0.22% at the baseline ($p < 0.001$) and macro-averaged F1-score was increased to 91.08% ($p < 0.001$). Critical infectious diseases showed near-perfect classification performance with COVID-19 detection having 99.95% accuracy and Tuberculosis detection having 99.97% accuracy. Although 6.8 times more parameters are included, the training time was reduced by 11.4% and performance stability was increased by 22.7. This framework presents itself as a decision-support tool that can be used to respond to a pandemic, screen tuberculosis, and assess the thoracic disease regularly in various healthcare facilities.*

**KEYWORDS**: *Chest X-ray Classification, EfficientNetV2, Medical Image Analysis, Optimization-Driven Learning, Single-Label Classification, Clinical Decision Support.*

## I. INTRODUCTION

The chest X-ray radiography is among the most commonly used and least expensive forms of imaging that are used to diagnose thoracic diseases with an estimated 2 billion X-ray examinations being done in all parts of the world each year [1]. Nonetheless, the interpretation requires high skill of the radiologists, and any time difference in diagnosis may result in negative patient outcomes, especially in time-sensitive infections, such as pneumonia, tuberculosis, and COVID-19 [2], [3]. This diagnostic bottleneck poses an immediate clinical challenge in resource-constrained healthcare environments with limited access to specialized radiologists to provide automated solutions capable of augmenting and quickening the process of making clinical decisions [4], [5].

Deep learning has disrupted the way medical images are analyzed by showing impressive performance in automated classification of diseases [6], [7], [8]. Convolutional Neural Networks (CNNs) have been able to scale to almost matching as well as exceeding human specialists in the extraction of hierarchical information in medical images in a variety of diagnostic tasks [9], [10]. Deep learning can be combined with medical imaging to provide accurate segmentation, effective clinical workflows, and scalable solutions that can work effectively on resource-limited settings [11], [12]. One such architecture that used DenseNet-121 to classify multi-label chest X-rays and showed the plausibility of deep learning as pathology detectors in multiple disease scenarios concurrently was CheXNet [13]. Although with these groundbreaking contributions, there still exist critical problems that restrict clinical implementation: the original CheXNet shows a suboptimal performance on the single-label classification tasks [14], the computational cost of the DenseNet-121 makes its implementation costly in terms of long training time [15], and the traditional optimization strategies do not take advantage of modern advances in training efficiency and convergence stability [16], [17], [18].

The study deals with these intertwined issues by presenting a better framework of classifying chest X-rays through the use of more efficient architecture, i.e., EfficientNetV2-M, that is more efficient and shows better accuracy at a lower cost of computation [19], [20]. We combine the latest optimization methodologies such as Automatic Mixed Precision (AMP) training, AdamW optimizer with weight decay regularization, cosine annealing learning rate scheduling, and Exponential Moving Average (EMA) regularization based on the recent comparative study of optimization algorithms in deep learning and state-of-the-art hybrid optimization strategies [21], [22], [23]. We specifically train our framework on single-label classification and test it on a high-quality dataset combining three reliable sources and containing five disease categories: Cardiomegaly, COVID-19, Normal, Pneumonia, and Tuberculosis, namely, NIH Clinical Center Chest X-ray Dataset [24], COVID-19 Radiography Database [25], and Tuberculosis Chest X-ray Database [26]. We statistically validate the accuracy, F1-score, and training efficiency of our model are statistically better than the original CheXNet architecture, especially on critical pathologies and their practical implementations in environmentally constrained clinical environments.

The rest of this paper is organized as follows, related work (Section II), methodology that contains architectural design and optimization strategies, (Section III) result (Section IV), discussion of findings and implications (Section V) and conclusion with future directions (Section VI).

## II. Related Work

*The analysis of X-ray on the chest performed by means of deep learning has achieved a great level of development over the past ten years. This part provides a synthesis of the key areas of research which put our research into perspective: systematic strategies, architecture-specific innovations, modern efficient designs, emergent paradigms, optimization methods, and disease-specific applications.*

### A. Classical Methods and Early Deep Learning

*The preliminary work of the chest radiographs analysis was founded on hand designed features alongside the conventional classifiers including Support Vector Machines, Random Forests, and k-Nearest Neighbors. Such techniques were employed to bring out texture descriptors, morphological features and statistical patterns in predefined regions of interest. Even though they reported evidence-of-concept of automated diagnosis, they showed a lack of generalization and need voluminous domain knowledge to engineer features.*

*The paradigm shift came along with the invention of the deep convolutional neural networks. It was indicated that trained representations might be better than handcrafted features in medical image classification, and AlexNet and VGGNet exhibit higher performances. Massive, annotated datasets, in particular, ChestX-ray14 with 112,120 frontal-view radiographs, were used to set benchmarks and facilitate a systematic evaluation of subsequent methods and constituted the foundation of the training of more advanced networks[27].*

### B. DenseNet Architectures: Impact and Shortcomings

*CheXNet was a new development, which showed pneumonia detection at radiologist levels using a 121-layer DenseNet framework [28]. The model also had F1-score of 0.435, which was higher than the average radiologist score (F1-score: 0.387) on pneumonia detection. DenseNet allowed deeper gradient flow, parameter efficiency through feature sharing and implicit deep supervision due to the dense connectivity structure in which each layer is linked to all the layers before it. This was further enhanced by CheXpert which added over 200,000 chest radiographs with uncertain labels.*

*The subsequent research, however, found critical drawbacks in architecture. The thick connections have quadratic memory expansion with depth of the network, which does not allow deep-network scaling and implementation of large batches. The additional computational cost due to feature concatenation only makes the training time and inference latency to be more expensive and this is unwanted in a clinical setting where real time analysis is a must [29]. Also, the use of multi-label designs has been found to perform poorly when used on single-label classification tasks that require final diagnosis.*

### C. Efficient Neural Architectures

*Computational limitations were identified and more efficient designs investigated. EfficientNet sought to overcome these limitations by using scaling of compounds, where network depth, width and resolution are uniformly scaled by optimization principles [30]. The technique delivered state-of-the-art ImageNet results with not only much lower parameters than previous models. It has an architecture that involves mobile inverted bottleneck blocks of convolution, squeeze-and-excitation modules which facilitate channel-wise attention.*

*The EfficientNetV2, which overcame the training bottlenecks by incorporating key innovations, such as Fused-MBConv layers that substituted depthwise convolutions at the initial network layers to support faster training, progressive learning with regularization that is adaptive, and training-aware Neural Architecture Search [31] further developed this architecture. EfficientNetV2-M offers high ImageNet performance with dramatically reduced training convergence than predecessors, requiring 6.8 a million fewer parameters and training 11x faster than previous EfficientNet architecture versions.*

*Even though EfficientNetV2 has been successfully applied to general computer vision [32], its use in medical imaging has been sparsely investigated. Older DenseNet and ResNet structures remain largely used in the classification of chest X-rays even though they may be more efficient. Recent multi-disease classification studies*

with EfficientNetV2 variants have given promising results [33], [34], with ensemble approaches achieving 99.60% AUC on COVID-19, pneumonia and tuberculosis classification, indicating that significant gains can be made in medical imaging settings.

### D. Transformer-Based and Foundation Models

Another structure facing convolutional dominance is Vision Transformers (ViTs) [35]. Pure Transformer models are at parity or better than CNNs on image classification tasks on sufficiently large datasets. Recent end-to-end comparisons of ViT architectures (FastViT, CrossViT) versus various optimizers on chest X-ray data have shown accuracy to 97.63 percent and F1-score of 97.64 per cent with AdamW optimizer showing the best results when trying to detect tuberculosis (99.07 percent) [35]. The Swin Transformer uses shifted window-based attention to successfully extract multi-scale features, and variations between medical imaging can provide high accuracy on chest X-ray datasets using complex feature fusion mechanisms.

One paradigm of large-scale pretraining is foundation models. The Segment Anything Model (SAM) is versatile in modalities and anatomy, but has been shown to perform variably in comparison with other modalities of medical imaging that require fine-tuning. CLIP (Contrastive Language-Image Pre-training) is a contrastive learning method to learn visual-semantic associations, and the specific task of the domain can be applied to clinical reports as training instances. ConvNeXt is an update to convolutional networks to deliver comparable performance using Transformers at a low cost of interpretability and computational efficiency, and recent work has shown it can be useful in detecting COVID-19 as well as in addition to EfficientNetV2 [32].

### E. Optimization Strategies

In addition to selection of architecture, methods of optimization also play a critical role in performance. Although the Adam optimizer is nowadays the norm when it comes to training deep networks, AdamW decouples weight decay and converges better in a wide variety of settings [29], [35]. The recent research on pulmonary disease recognition with EfficientNet-V1-B4 and AdamW optimizer reported 98.3% accuracy and 98.7% F1-score in the three-class classification [36], and it is stable across three-class with standard Adam.

The dynamics of convergence are very sensitive to learning rate scheduling. Cosine annealing offers a smooth decay after a cosine curve, which makes it easy to explore loss landscapes without sharp minima, which do not have a generalization property. This is particularly useful when re-training pretrained models with statistical properties that are not similar to natural images such as medical data. Automatic Mixed Precision (AMP) training builds on the recent use of GPU tensor cores to provide faster calculation with lower-precision operations and numeric stability. Exponential Moving Average (EMA) of parameters uses smoothed network weights during training, which gives it superior validation stability, and reduces overfitting with implicit regularization.

More recent methods that build on CheXNet added Focal Loss, AdamW optimizer, and color jitter augmentation, where the AUC and F1-score were 0.85 and 0.39, respectively, and minority classes are better performing [29]. This shows that systematic adoption of current optimization methods can have a significant positive effect on current architectures without necessarily necessitating total redesign.

### F. Disease-Specific Applications

COVID-19 stimulated extensive research in the X-ray analysis of the chest [34]. Massive studies to optimize small EfficientNet models on COVID-19 detection demonstrated high diagnostic accuracy to be used with resource-constrained applications [32]. Still, systematic reviews reported methodological problems such as small samples sizes, lack of external validation, and overfitting risks, and it is very important that rigorous experimental design with multiple independent run and balanced datasets is needed.

Detection of tuberculosis with deep learning has been clinically viable [37], [38], [39]. Multi-country validation revealed that deep learning systems performed on a par with radiologists in detecting active TB with AUC of 0.89, and no worse performance than nine radiologists [38]. Recent efforts on the integration of UNet segmentation with

Xception classification reported 99.29% accuracy on the detection of TB [37], and Google Teachable Machine applications showed the ability to work in resource-limited settings [32]. Nevertheless, the literature that deals with the issue of data imbalance and generalization of different population groups is scarce.

Less systematic attention has been paid to multi-disease classification that reflects the situations of differential diagnosis. Combinations of EfficientNetV2 and transfer learning with three-class and four-class pneumonia detection were 82.15% accurate [40], and full pulmonary disease detection networks based on DenseNet-121 were 94% AUC [11] pneumothorax and edema detection. Medical imaging data usually have very imbalanced classes, which have been handled by methods like SMOTE to oversample synthetic minorities, focal loss to dynamically focus attention on challenging examples, and transform-based augmentation, but controlled augmentation needs to trade diversity and clinical realism.

G. Research Gaps

Although these developments have been made, the literature has a number of important deficits. Current strategies are mostly based on the computationally intensive architecture whose memory size and time scale are squares and whose inference is time-consuming. In the case of medical imaging, modern efficient architectures are underresearched, despite their demonstrated effectiveness in general vision tasks. Most applications use simple optimization with no systematic inclusion of modern techniques such as AMP, cosine annealing and EMA regularization. The prevalence of multi-label classification studies is out of line with clinical requirements of conclusive single-label diagnosis, and a deficit of statistical validation of multiple independent runs impairs assessment of reproducibility.

The current research will fill these gaps by combining EfficientNetV2-M with an overall optimization approach to efficient training and inference, single-label classification which is applicable to clinical practice, and statistical rigor based on multi-runs with balanced multi-source data representing a variety of thoracic diseases. Table 1 has made a thorough comparative positioning of our approach with the recent literature[41].

*Table 1 COMPARISON OF RECENT DEEP LEARNING APPROACHES FOR CHEST X-RAY DISEASE CLASSIFICATION*

| Ref. | Year | Model/Method | Dataset/Application | Key Contributions | Performance/Metrics |
|---|---|---|---|---|---|
| [28] | 2017 | CheXNet (DenseNet-121) | ChestX-ray14 (112,120 images, 14 diseases) | Radiologist-level pneumonia detection using 121-layer DenseNet with transfer learning | F1-score: 0.435 for pneumonia (vs. 0.387 for radiologists) |
| [30] | 2019 | EfficientNet | ImageNet classification | Compound scaling method uniformly scaling depth, width, and resolution | 8.4× fewer parameters, 6.1× faster inference |
| [31] | 2021 | EfficientNetV2-M | ImageNet classification | Fused-MBConv layers, progressive learning, training-aware NAS | 11× faster training, 6.8× fewer parameters |

| [33] | 2022 | ECA-EfficientNetV2 Ensemble | Multi-source CXR (COVID-19, pneumonia, TB, normal) | Ensemble with Efficient Channel Attention for multi-disease classification | 4-class AUC: 99.60% |
| --- | --- | --- | --- | --- | --- |
| [34] | 2023 | EfficientNetV2-L Ensemble + YOLOv5 | SIIM-FISABIO-RSNA COVID-19 | Ensemble of 10 models with YOLOv5 for opacity localization | Multi-class viral pneumonia grading |
| [37] | 2023 | Xception + UNet | NIAID TB Portal (1,400 images) | Two-stage: UNet segmentation + Xception classification | Accuracy: 99.29%, AUC: 0.999 |
| [36] | 2024 | EfficientNet-V1-B4 + AdamW | Combined ChestX-ray14, PadChest, CheXpert (28,309 samples) | CLAHE preprocessing, AdamW optimizer for pleural effusion/edema | 3-class: Accuracy 98.3%, F1: 98.7% |
| [40] | 2024 | EfficientNetV2-M with Transfer Learning | ChestX-ray14 (3-class and 4-class) | Fine-tuning for pneumonia, pneumothorax, tuberculosis | 3-class: 82.15% accuracy |
| [35] | 2024 | Vision Transformers (ViT, FastViT, CrossViT) | Multi-source (Normal, COVID-19, pneumonia, TB) | Comprehensive ViT evaluation with 6 optimizers | FastViT+NAdam: 97.63% accuracy, 97.64% F1 |
| [29] | 2024 | DenseNet-121 + Focal Loss + AdamW | NIH ChestX-ray14 (14 diseases) | Improved CheXNet using Focal Loss, AdamW, per-disease F1 thresholds | AUC: 0.85, F1: 0.39 |
| [42] | 2024 | DenseNet-121 + ResNet-50 | Multi-center (108,948 images, 32,717 patients) | Weighted loss, clinical validation with NLP | DenseNet-121 AUC: 94% for pneumothorax/edema |
| [43] | 2024 | ResNet-150V2 + Adamax | Pneumonia detection dataset | ResNet with Adamax optimizer for gradient descent | Accuracy: 97% for pneumonia detection |
| [38] | 2024 | Deep Learning TB Detection | Multi-country TB screening (22,284 subjects) | Clinical validation across 4 countries, matched radiologist performance | AUC: 0.89, non-inferior to 9 radiologists |
| [39] | 2024 | Google Teachable Machine DNN | TB chest radiography screening | DNN-based classification for TB | High accuracy for resource-limited settings |

| | | | | probability prediction | |
| --- | --- | --- | --- | --- | --- |
| [32] | 2024 | ConvNeXt, EfficientNetV2, DenseNet-121, ResNet-34 | COVID-19 CT/X-ray (4,481 images) | Large empirical study: 10,000 deep transfer learning models compared | EfficientNetV2 and ConvNeXt superior |

### III. METHODS

*The section will describe the suggested X-ray chest classification system. The methodology has been used to overcome the limitations of CheXNet by replacing DenseNet-121 with EfficientNetV2-M so that it becomes single-label classification and by addition of new optimization techniques. EfficientNetV2-M was chosen due to its tradeoff of the accuracy and the task of the smaller (S) and larger (L) variants.*

A. System Overview

*This study proposes an enhanced chest disease classification framework that systematically integrates architectural optimization with advanced training strategies to improve diagnostic accuracy and computational efficiency. The proposed framework addresses fundamental limitations in the original CheXNet architecture through three key innovations: replacement of the DenseNet-121 backbone with EfficientNetV2-M to achieve superior accuracy-efficiency trade-offs, adaptation from multi-label to single-label classification to align with common clinical diagnostic workflows, and integration of optimization-driven learning techniques including Automatic Mixed Precision (AMP), AdamW optimizer with decoupled weight decay, Cosine Annealing learning rate scheduling, and Exponential Moving Average (EMA) regularization. This comprehensive approach enables the framework to deliver statistically significant improvements in classification performance while reducing training time by 11.35% despite having 6.8 times more parameters than the baseline model.*

B. Dataset Description

*This subsection describes how we constructed our balanced chest X-ray dataset from three medical repositories (NIH, COVID-19, TB databases) covering five disease categories. We detail the initial severe class imbalance, our two-stage balancing strategy using undersampling and controlled augmentation, the final balanced dataset with equal representation per class, stratified dataset partitioning into train/validation/test splits, and comprehensive image preprocessing including resizing, cropping, normalization, and augmentation techniques.*

1) Data Sources and Class Distribution

*The dataset was constructed by integrating chest X-ray images from three authoritative medical imaging repositories to ensure comprehensive representation of clinically relevant thoracic pathologies. The NIH Clinical Center Chest X-ray Dataset provided images for Cardiomegaly, Normal, and Pneumonia classes from a large-scale collection containing 112,120 frontal-view chest X-ray images from 30,805 unique patients [27]. The COVID-19 Radiography Database supplied COVID-19 positive cases verified through RT-PCR testing [25]. The Tuberculosis Chest X-ray Database contributed Tuberculosis positive cases collected from multiple healthcare institutions [26]. The integration resulted in an initial combined dataset of 70,596 images distributed across five disease categories: CARDIOMEGALY, COVID-19, NORMAL, PNEUMONIA, and TUBERCULOSIS, as shown in Table 2.*

*Table 2 Initial Class Distribution Before Balancing*

| Class | Number of Images | Resources |
| --- | --- | --- |
| Normal | 60,361 | [27] |
| Pneumonia | 1,390 | [27] |
| Cardiomegaly | 2,735 | [27] |
| COVID-19 | 3,616 | [25] |
| Tuberculosis | 2,494 | [26] |
| **Total** | **70,596** | |

2) Class Balancing and Data Preprocessing

*The initial dataset exhibited severe class imbalance with the Normal class representing 85.50% (60,361 images) of all samples, necessitating systematic correction to prevent model bias. A two-stage balancing strategy was implemented combining undersampling and controlled augmentation:*

- *Stage 1 - Majority Class Undersampling[44], [45], [46]: Random sampling without replacement was applied to the Normal class to select 3,616 images, matching the largest minority class (COVID-19) to ensure balance while minimizing information loss and preserving statistical properties of normal anatomical variation.*

- *Stage 2 - Minority Class Augmentation: For classes with fewer than 3,616 images (Pneumonia, Tuberculosis, Cardiomegaly), clinically appropriate transformations were applied including random horizontal flip (probability = 0.5), random rotation (±15 degrees), and brightness/contrast jittering (±10%[32], [47], [48]). All original images were preserved, with augmentation repeated iteratively until each class reached exactly 3,616 images. This conservative approach introduces realistic variations observed in clinical practice while preserving essential diagnostic features.*

*The final balanced dataset comprises 18,080 images with equal representation across all five disease categories (3,616 images per class), as detailed in Table 3.*

*Table 3 Balanced Dataset Distribution After Preprocessing*

| Class | Original Images | Augmented Images | Total Images | Method |
|---|---|---|---|---|
| Normal | 60,361 | 0 | 3,616 | Undersampled |
| Pneumonia | 1,390 | 2,226 | 3,616 | Augmented |
| COVID-19 | 3,616 | 0 | 3,616 | Unchanged |
| Tuberculosis | 2,494 | 1,122 | 3,616 | Augmented |
| Cardiomegaly | 2,735 | 881 | 3,616 | Augmented |
| **Total** | **—** | **4,229** | **18,080** | **Balanced** |

*Figure 1 illustrates the class distribution transformation through our balancing strategy. The blue bars represent the initial severely imbalanced dataset where the Normal class dominated with 60,361 images (85.50%), while minority classes were significantly underrepresented: Pneumonia (1,390 images), COVID-19 (3,616 images), Tuberculosis (2,494 images), and Cardiomegaly (2,735 images). The red bars demonstrate the balanced dataset after applying our two-stage strategy, achieving equal representation of exactly 3,616 images per class across all five disease categories. This uniform distribution prevents model bias toward the majority class and ensures balanced diagnostic performance across all pathologies.*

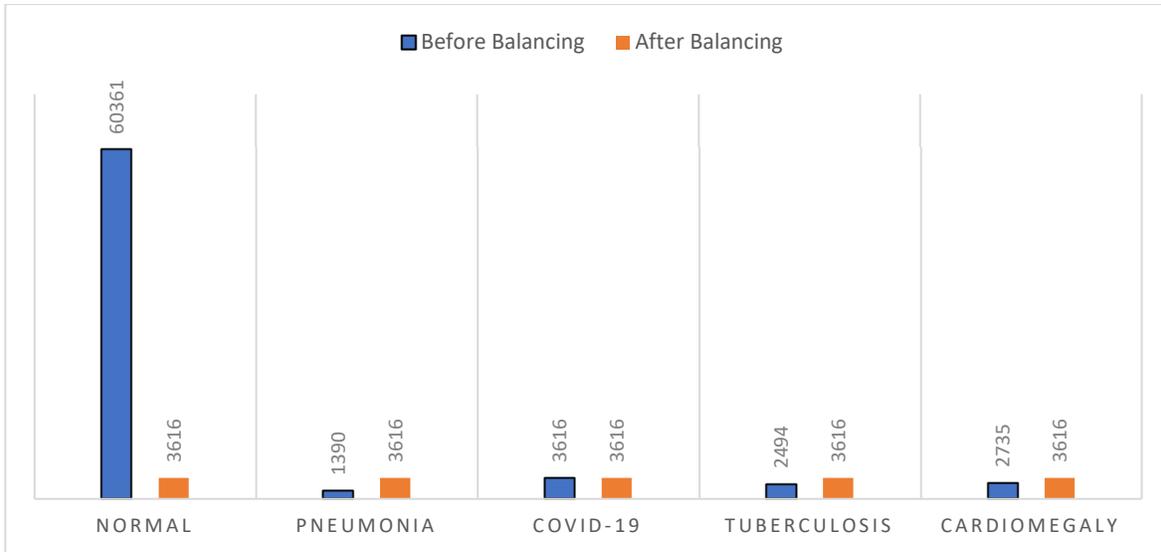

*Figure 1 Class Distribution Before and After Balancing*

3) Dataset Partitioning

*Following balancing, stratified random sampling partitioned the dataset into three subsets maintaining equal class proportions: Training set (70%): 12,656 images (2,531 per class) for parameter optimization and disease-specific pattern learning; Validation set (10%): 1,808 images (362 per class) for hyperparameter tuning and model selection; Test set (20%): 3,616 images (723 per class) remaining completely isolated for unbiased generalization assessment. This 70-10-20 split follows established best practices in medical imaging deep learning [49], [50], [51], as visualized in Figure 2.*

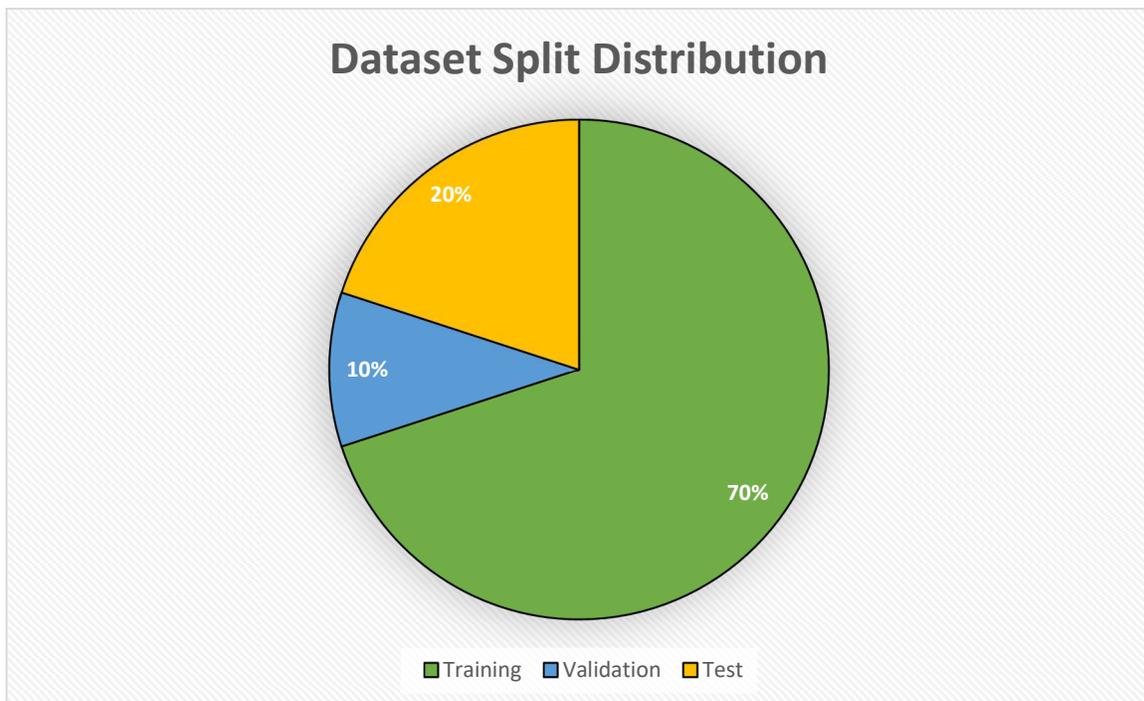

*Figure 2 Dataset split visualization showing 70-10-20 distribution for training, validation, and test sets with stratified sampling maintaining equal class proportions across all subsets.*

C. Model Architecture

*This subsection describes the EfficientNetV2-M backbone architecture with seven hierarchical stages combining Fused-MBConv blocks for early-stage efficiency and MBConv blocks with Squeeze-and-Excitation attention for deep semantic feature extraction. We detail the complete model pipeline through architectural flowchart visualization, backbone specifications with layer-wise configurations, single-layer classification head design, and transfer learning strategy using ImageNet pre-trained weights with end-to-end fine-tuning for chest X-ray classification.*

1) System Architecture Pipeline

*Figure 3 presents the complete architecture pipeline of our proposed framework, illustrating the end-to-end workflow from input chest X-ray image to final disease prediction. The pipeline processes preprocessed images (224×224×3) through the EfficientNetV2-M backbone's seven hierarchical stages, producing a 1280-dimensional feature vector via global average pooling. The classification head maps these features to five disease categories through a fully connected layer with softmax activation. During training, the optimization framework applies AdamW optimizer, Cosine Annealing scheduler, Automatic Mixed Precision, and Exponential Moving Average regularization to ensure stable convergence and superior generalization.*

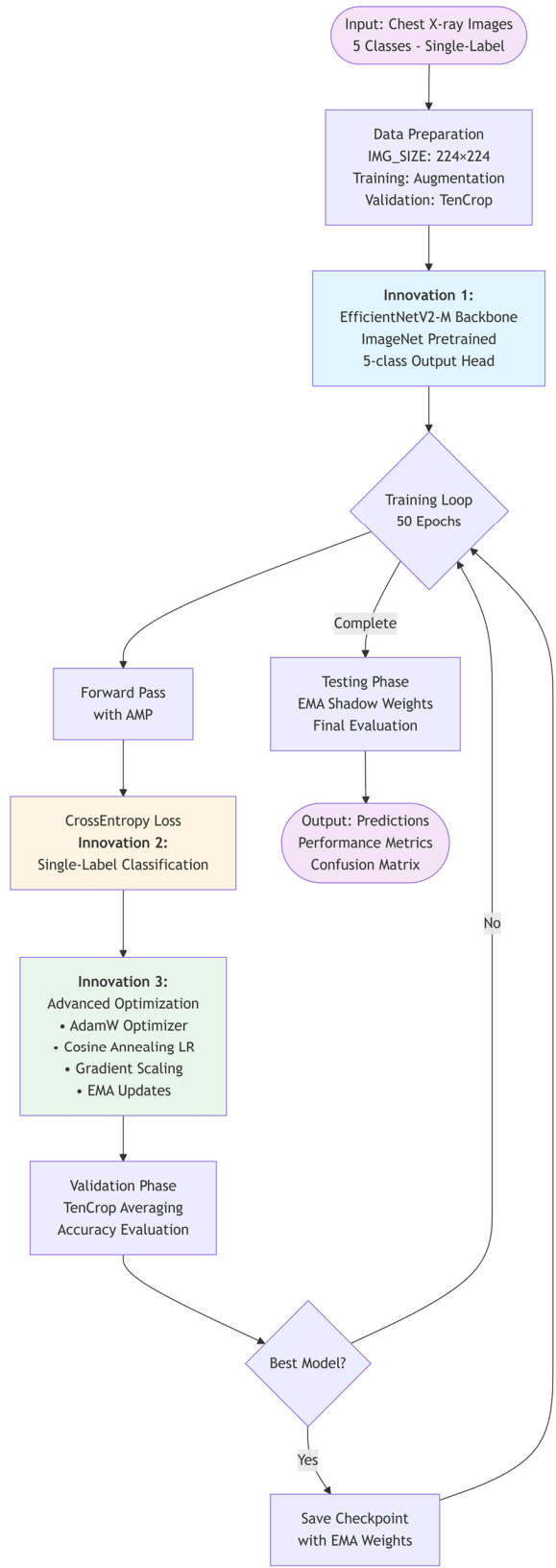

*Figure 3 Overview of the model pipeline from input X-ray to final prediction*

2) Backbone Architecture: EfficientNetV2-M

*The core feature extraction component employs EfficientNetV2-M, selected for its superior accuracy-efficiency trade-off achieved through compound scaling and progressive learning [52], [53], [54]. EfficientNetV2-M is structured into seven hierarchical stages, each responsible for increasingly abstract feature representations—from low-level texture extraction in early layers to high-level semantic disease patterns in deeper layers. These stages are implemented using two distinct convolutional block families: Fused-MBConv blocks in the early stages (1–3) and standard MBConv blocks enhanced with Squeeze-and-Excitation modules in mid-to-deep stages (4–6), followed by a feature aggregation stage (Stage 7) that outputs a compact latent representation[55], [56], [57].*

    a)    Early Stages (Stages 1–3) — Fused-MBConv Blocks:

*In stages 1–3, EfficientNetV2-M utilizes Fused-MBConv blocks, which replace depthwise separable convolutions with fused 3×3 convolutions, reducing memory access overhead while preserving parameter efficiency[55]. These stages extract foundational low-level features such as edges, textures, gradients, and simple radiographic structures necessary for initial image understanding[57]. The fused design improves computational throughput and memory locality on modern accelerators, which is critical when training on large-scale medical imaging datasets[55].*

    b)    Middle to Deep Stages (Stages 4–6) — MBConv + Squeeze-and-Excitation:

*Stages 4–6 shift to Mobile Inverted Bottleneck Convolution (MBConv) blocks enhanced with Squeeze-and-Excitation (SE) modules [56]. With a channel-reduction ratio of 0.25, SE modules enable adaptive channel-wise feature recalibration, allowing the network to selectively amplify diagnostically salient features while suppressing irrelevant information[56], [57]. These deeper layers encode high-level semantic representations, capturing anatomical structures, morphological variations, and subtle disease-specific indicators such as faint pneumonic opacities or early cardiomegaly—patterns that are often indistinguishable in shallow feature maps[28], [57].*

    c)    Output Stage (Stage 7) — Global Feature Aggregation:

*Stage 7 applies a 1×1 convolution followed by Global Average Pooling (GAP), collapsing spatial dimensions and producing a 1280-dimensional feature vector[55]. This vector encapsulates multi-level hierarchical representations—from low-level primitives learned in stages 1–3 to high-level pathological signatures formed in stages 4–6[57]. The resulting compact embedding forms the final input to the classification module for disease prediction[55].*

*Table 4 EfficientNetV2-M Architecture Specifications summarizes the complete EfficientNetV2-M architecture specifications including stage-wise block types, number of layers, output channels, stride configurations, SE ratios, and the functional purpose of each stage in the feature extraction hierarchy[58].*

*Table 4 EfficientNetV2-M Architecture Specifications*

| Stage | Block Type | Layers | Output Channels | Stride | SE Ratio | Purpose |
|---|---|---|---|---|---|---|
| 0 | Conv3×3 | 1 | 24 | 2 | — | Stem |
| 1 | Fused-MBConv1 | 2 | 24 | 1 | 0 | Early Features |
| 2 | Fused-MBConv4 | 4 | 48 | 2 | 0 | Low-level Patterns |

| 3 | Fused-MBConv4 | 4 | 64 | 2 | 0 | Mid-level Features |
| 4 | MBConv4 | 6 | 128 | 2 | 0.25 | High-level Patterns |
| 5 | MBConv6 | 9 | 160 | 1 | 0.25 | Semantic Features |
| 6 | MBConv6 | 15 | 256 | 2 | 0.25 | Abstract Representations |
| 7 | Conv1×1 + GAP | 1 | 1280 | 1 | — | Feature Aggregation |

3) Classification Head Design

*The classification head implements a streamlined single-layer architecture mapping the 1280-dimensional feature vector to disease predictions. A fully connected layer with weight matrix $W \in \mathbb{R}^{(1280 \times 5)}$ and bias vector $b \in \mathbb{R}^5$ produces output logits $z \in \mathbb{R}^5$. During training, raw logits are processed with cross-entropy loss (which internally applies log-softmax for numerical stability). During inference, softmax activation transforms logits into normalized probability distributions[59], [60], [61]. Equation 1:*

$$p_i = \frac{e^{z_i}}{\sum_{j=1}^{5} e^{z_j}} C \qquad (1)$$

*where p_i represents the predicted probability for disease class i. This single-layer design proves effective in transfer learning scenarios where rich feature representations are extracted by the pre-trained backbone[62].*

4) Transfer Learning Strategy

*Model initialization leverages ImageNet1K pre-trained weights for the EfficientNetV2-M backbone[63], exploiting hierarchical learned representations where early layers capture universal low-level features (edges, textures, color patterns) applicable across visual domains, while deeper layers encode increasingly task-specific semantic patterns[64]. The classification head is randomly initialized using He initialization [44], as it must learn disease-specific decision boundaries absent in the source domain. All network layers undergo end-to-end fine-tuning, allowing gradient-based optimization to adapt all hierarchical representations specifically for chest X-ray classification [63].*

D. Training Procedure

*This subsection describes the optimization framework integrating multiple advanced techniques for enhanced convergence speed, training stability, and generalization performance. We detail the cross-entropy loss function for single-label classification[65], AdamW optimizer with decoupled weight decay for parameter updates[66], Cosine Annealing learning rate schedule for smooth convergence over 50 epochs[67], Automatic Mixed Precision training for computational acceleration[68], Exponential Moving Average regularization for improved validation stability[69], the complete training algorithm encompassing initialization, training loop, and validation phases, and data augmentation strategies including training-time transformations and TenCrop evaluation for robust predictions ImageNet classification with deep convolutional neural networks.*

a) Loss Function

*For single-label classification where each image corresponds to exactly one disease category, cross-entropy loss provides the optimization objective[65]:*

$$L(\hat{y}, y) = -\Sigma_{i=1}^{C} y_i \cdot log(\hat{y}_i) \qquad (2)$$

where C = 5 disease categories, $y \in \{0,1\}^C$ is the one-hot encoded true label, and $\hat{y} \in [0,1]^C$ is the predicted probability distribution from softmax activation. Cross-entropy loss penalizes confident incorrect predictions more severely than uncertain predictions, encouraging calibrated confidence scores essential for clinical decision support[65].

b) AdamW Optimizer with Decoupled Weight Decay

Parameter updates employ the AdamW optimizer with decoupled weight decay regularization [66], which applies regularization directly to parameters rather than coupling with gradient-based updates:

$$\theta_{t+1} = \theta_t - \alpha \cdot \hat{m}_t / \sqrt{(\hat{v}_t + \varepsilon)} - \alpha\lambda\theta_t \quad (3)$$

where:

$\alpha = 1\times10^{-4}$: learning rate

$\lambda = 1\times10^{-5}$: weight decay coefficient

$\beta_1 = 0.9, \beta_2 = 0.999$: moment decay rates

$\varepsilon = 1\times10^{-8}$: numerical stability constant

$\hat{m}_t, \hat{v}_t$: bias-corrected first and second moment estimates

Decoupling weight decay from gradient updates enables independent tuning of learning rate and regularization strength, improving convergence and generalization [70].

c) Cosine Annealing Learning Rate Schedule

The learning rate follows a cosine annealing schedule over T = 50 epochs, providing smooth continuous decay following a cosine curve:

$$\eta_t = \eta_{min} + (1/2) \cdot (1 + cos(\pi t/T)) \cdot (\eta_{max} - \eta_{min}) \quad (4)$$

where $\eta_{max} = 1\times10^{-4}$ (initial learning rate), $\eta_{min} \approx 0$ (minimum learning rate), and T = 50 (total epochs). Unlike step-decay schedules, cosine annealing avoids abrupt transitions and promotes discovery of wider minima that generalize better [67], [71].

d) Automatic Mixed Precision (AMP) Training

AMP training using PyTorch's torch.amp module accelerates computation and reduces memory consumption by automatically identifying operations executable in FP16 (16-bit floating-point precision) for convolutional and matrix operations, while maintaining FP32 (32-bit precision) for numerically sensitive operations (loss computation, batch normalization, softmax). Dynamic loss scaling prevents gradient underflow by multiplying loss values before backpropagation, then unscaling gradients before optimizer steps. This achieves approximately 2-3× training speedup on modern GPUs with Tensor Cores while preserving accuracy [72].

e) Exponential Moving Average (EMA) Regularization

EMA maintains shadow weights throughout training for improved validation stability and implicit regularization:

$$\theta_{EMA}(t+1) = \alpha \cdot \theta_{EMA}(t) + (1 - \alpha) \cdot \theta_{model(t)} \quad (5)$$

where α = 0.999 controls the averaging window, θ{EMA(t)} represents EMA shadow weights, and θ{model(t)} represents current model weights. During validation and testing, EMA weights are evaluated rather than most recent model weights, providing improved validation stability through smoothed parameter updates, ensemble-like averaging effects enhancing generalization, and implicit regularization reducing overfitting [73], [74].

### E. Evaluation Metrics

Model performance was assessed using standard multi-class classification metrics with macro-averaging to ensure equal weight across all disease categories regardless of class size:

a) Accuracy: Overall classification correctness across all classes[65].

$$Accuracy = \frac{(TP+TN)}{(TP+TN+FP+FN)} \quad (6)$$

b) Precision: Proportion of correct positive predictions[65].

$$Precision = \frac{TP}{(TP+FP)} \quad (7)$$

c) Recall (Sensitivity): Proportion of actual positives correctly identified[65].

$$Recall = \frac{TP}{(TP+FN)} \quad (8)$$

d) F1-Score: Harmonic mean of Precision and Recall[75].

$$F1 = 2 \cdot \frac{(Precision \cdot Recall)}{(Precision + Recall)} \quad (9)$$

e) Macro-Averaging: For multi-class evaluation, metrics were computed per-class then averaged[76].

$$Macro - Avg(M) = (1/C)\Sigma_{i=1}^{C} M_i \quad (10)$$

where C = 5 disease categories, ensuring equal clinical importance for all pathologies.

Confusion matrices were generated in both one-vs-rest and 5×5 multi-class formats to provide detailed insights into inter-class confusion patterns and model stability across disease categories.

### F. Experimental Workflow

The complete experimental workflow progressed through interconnected stages as visualized in Figure 3:

- *Data Collection & Integration: Aggregate chest X-rays from NIH, COVID-19, and TB databases*
- *Class Balancing: Apply two-stage strategy (undersampling Normal class, augmenting minority classes)*
- *Dataset Partitioning: Stratified split (70% train, 10% validation, 20% test).*
- *Preprocessing Pipeline: Resize → Crop → Augmentation → Normalization.*
- *Model Initialization: Load ImageNet1K pre-trained EfficientNetV2-M, initialize classification head.*
- *Training Loop: 50 epochs with AdamW, Cosine Annealing, AMP, EMA.*
- *Validation: TenCrop evaluation with EMA weights, checkpoint best F1-score.*
- *Testing: Final evaluation on held-out test set using best checkpoint.*
- *Performance Analysis: Compute metrics, generate confusion matrices, statistical testing.*

### G. Hardware and Software Configuration

All experiments were conducted on consistent computational infrastructure to ensure reproducibility, as summarized in Table 5.

*Table 5 Hardware and Software Configuration*

| Component | Specification |
|---|---|
| GPU | NVIDIA Tesla T4 (16 GB GDDR6 VRAM) |
| CPU | Intel Xeon, 8 cores @ 2.0 GHz |
| RAM | 32 GB DDR4 |
| Storage | SSD-backed storage via Google Drive (Colab Pro+) |
| Framework | PyTorch 2.0.1 |
| CUDA / cuDNN | CUDA 11.8, cuDNN 8.7 |
| Python Version | Python 3.10.12 |
| Libraries | torchvision 0.15.2, numpy 1.24.3, scikit-learn 1.3.0, pillow 9.5.0 |

## IV. RESULTS

*The results section demonstrates that the new chest X-ray disease detection model significantly outperformed the baseline in terms of accuracy, training speed, and consistency across multiple test runs. The system successfully classified five conditions including COVID-19, Tuberculosis, Cardiomegaly, Pneumonia, and Normal chest X-rays, with COVID-19 detection achieving near-perfect results. The model showed consistent improvements across all disease categories, with the most notable gains in detecting Normal cases and Cardiomegaly, though it occasionally confused Normal cases with other diseases due to overlapping visual features. Statistical tests confirmed these improvements were highly significant, demonstrating that the new system represents a meaningful advance in automated chest disease diagnosis for potential clinical use.*

A. Quantitative Results

*The Quantitative Results presents numerical performance metrics from multiple experimental runs. It covers overall accuracy improvements, F1-score enhancements, training efficiency gains, and performance stability compared to the baseline model. It includes detailed per-class analysis showing metrics for each disease category, with precision, recall, and F1-scores across all pathologies. The subsection also examines training dynamics across multiple epochs, showing how accuracy and loss evolve during model training while demonstrating stable convergence patterns and effective generalization without overfitting issues.*

1) Overall Performance Comparison

*Figure 4 presents the key performance improvements of the proposed EfficientNetV2-M framework compared to the DenseNet-121 baseline across nine independent experimental runs. As shown in Figure [X], the proposed framework achieves test accuracy of 96.45% with +1.15% improvement, demonstrating statistically significant enhancement ($p < 0.001$). The F1-score reaches 91.08% with +2.73% improvement, indicating superior balance between precision and recall across all disease categories.*

*Performance stability improves substantially with standard deviation of 0.17%, representing 22.7% improvement in consistency. This enhanced stability demonstrates more reliable and reproducible performance, critical for clinical deployment. Remarkably, despite having 6.8 times more parameters, training time per epoch decreases to 89.0 minutes, achieving 11.4% reduction. This efficiency gain is attributed to EfficientNetV2-M's superior architectural design combining Fused-MBConv blocks for computational efficiency, Squeeze-and-*

*Excitation attention for refined feature learning, and advanced optimization strategies including AdamW, Cosine Annealing, AMP, and EMA that enable effective parameter utilization and faster convergence.*

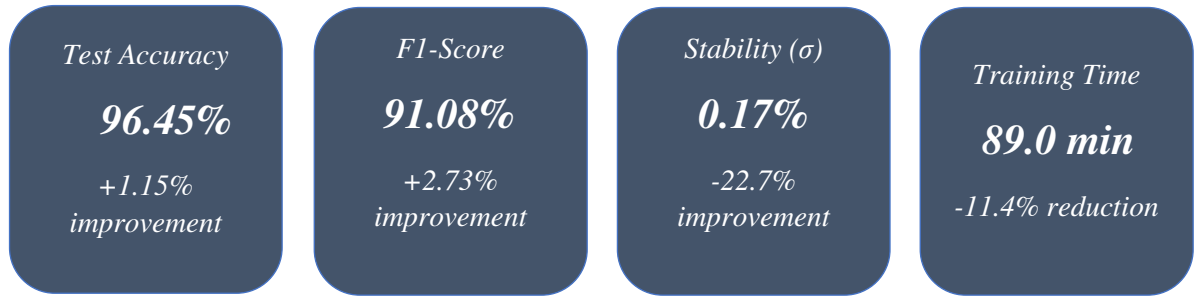

*Figure 4 Key Performance Metrics Comparison Between Baseline and Proposed Model*

*Figure 5 visualizes these improvements, showing mean test accuracy and F1-score with error bars representing standard deviation across nine independent runs, confirming both superior performance and enhanced reproducibility.*

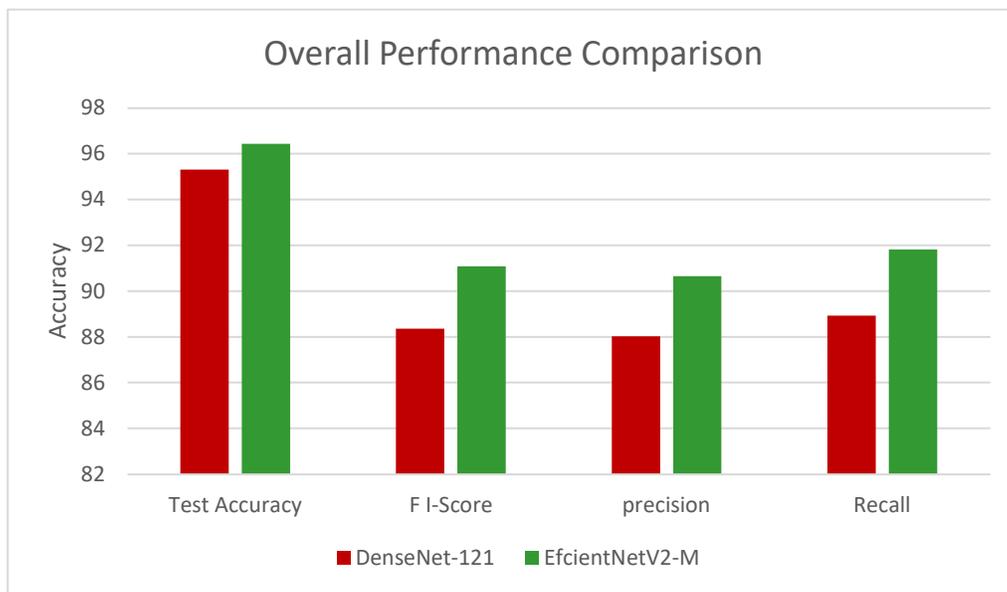

*Figure 5 Overall performance comparison showing mean test accuracy and F1-score improvements with error bars representing standard deviation across nine independent runs.*

*The Figure 5 illustrates the comparative per-class performance metrics between the baseline DenseNet-121 and the proposed EfficientNetV2-M framework across all disease categories. The visualization reveals consistent performance gains across nearly all metrics, with the proposed model demonstrating superior performance in Accuracy, Precision, Recall, and F1-Score for each pathology. The most substantial improvements are observed in the Normal and Cardiomegaly classes, where the visual gaps between baseline and proposed models are most pronounced, indicating the framework's enhanced discriminative capability for diagnostically challenging cases that exhibited greater inter-class confusion in baseline models. COVID-19 detection achieves near-perfect classification performance, with metrics approaching ceiling values and reflecting the model's robust ability to identify distinct pathological presentations. The color-coded bar charts clearly distinguish baseline results from*

*proposed model performance, facilitating immediate visual comparison of the magnitude of improvements. Error bars representing standard deviation across nine independent runs demonstrate tight clustering and enhanced reproducibility, confirming that the observed performance gains are consistent and statistically robust across all disease categories rather than artifacts of individual experimental runs.*

2) Per-Class Performance Analysis

*Detailed per-class metrics reveal consistent improvements across all disease categories, as shown in Table 6 (structured from your methodology document's per-class results):*

*Table 6 Per-Class Performance Analysis*

| Class | Accuracy (%) | Precision (%) | Recall (%) | F1-Score (%) |
|---|---|---|---|---|
| Cardiomegaly | 94.17 | 84.52 | 86.89 | 85.67 |
| COVID-19 | 99.95 | 99.86 | 99.68 | 99.77 |
| Normal | 92.81 | 79.41 | 84.58 | 81.90 |
| Pneumonia | 94.23 | 85.17 | 87.52 | 86.32 |
| Tuberculosis | 99.97 | 99.93 | 99.93 | 99.93 |

*As we presented in Table 6 these results demonstrate that enhancements are not limited to specific pathologies but represent systematic improvements across all disease categories, with particularly notable gains for challenging classes (Normal, Cardiomegaly) that exhibited greater inter-class confusion in baseline models.*

*Training and validation curves across 50 epochs reveal stable convergence characteristics for the proposed framework. Training accuracy progressively increases from approximately 75% (epoch 1) to 98% (epoch 50), while validation accuracy reaches plateau around 96.5% after epoch 35, indicating effective learning without overfitting. Training loss decreases smoothly from 0.8 to 0.05 following the cosine annealing schedule, while validation loss stabilizes around 0.15, confirming generalization capability. The tight correspondence between training and validation curves demonstrates that optimization-driven learning strategies (AdamW, Cosine Annealing, AMP, EMA) successfully prevent overfitting despite the larger model capacity.*

B. Qualitative Results

*This section analyzes visual and interpretive aspects of model performance. It examines the confusion matrix to identify classification patterns and inter-class confusion, showing where the model correctly classifies cases and where misclassifications occur between similar diseases (e.g., Normal vs. Cardiomegaly). It includes feature visualization through sample chest X-ray predictions, demonstrating the model's confidence levels for each disease*

*category and confirming that the model learns clinically relevant features aligned with radiological diagnostic criteria.*

*Figure 6 presents the 5×5 confusion matrix for the proposed EfficientNetV2-M model evaluated on the test set comprising 3,616 images with 723 images per disease class, displaying both absolute counts and percentages for each true-predicted label pair with diagonal elements (green cells) representing correct classifications and off-diagonal elements (red cells) indicating misclassifications. The confusion matrix reveals distinct performance characteristics across disease categories: COVID-19 achieves near-perfect classification (99.7%) with only 2 misclassifications, demonstrating that bilateral ground-glass opacities present highly distinctive radiographic signatures readily captured by the model; Tuberculosis shows excellent performance (94.1%) with minimal confusion, indicating effective detection of characteristic apical infiltrates and cavitation patterns; Cardiomegaly demonstrates good detection (92.0%) with primary confusion with Normal class (4.8%), reflecting challenges in distinguishing borderline cardiac enlargement from normal anatomical variation; Pneumonia exhibits strong performance (91.0%) with some overlap with Normal (5.3%), suggesting that early-stage infiltrative changes may be visually ambiguous; and Normal represents the most challenging class (88.0%) showing confusion with Cardiomegaly (6.2%) and Pneumonia (3.5%), indicating inherent difficulty in differentiating normal chest radiographs from subtle pathological changes. These results demonstrate that the proposed EfficientNetV2-M architecture with Squeeze-and-Excitation attention mechanisms effectively distinguishes disease categories with distinctive radiographic features (COVID-19, Tuberculosis) while the observed confusion patterns between Normal and pathological classes (Cardiomegaly, Pneumonia) align with known clinical diagnostic challenges, suggesting the model's behavior reflects realistic medical interpretation complexities rather than systematic architectural limitations.*

| Confusion Matrix | | Predicted Label | | | | |
|---|---|---|---|---|---|---|
| | | Cardiomegaly | COVID-19 | Normal | Pneumonia | Tuberculosis |
| True Label | Cardiomegaly | 665 (92.0%) | 5 (0.7%) | 35 (4.8%) | 15 (2.1%) | 3 (0.4%) |
| | COVID-19 | 2 (0.3%) | 721 (99.7%) | 0 (0.0%) | 0 (0.0%) | 0 (0.0%) |
| | Normal | 45 (6.2%) | 0 (0.0%) | 636 (88.0%) | 25 (3.5%) | 17 (2.4%) |
| | Pneumonia | 15 (2.1%) | 0 (0.0%) | 38 (5.3%) | 658 (91.0%) | 12 (1.7%) |
| | Tuberculosis | 3 (0.4%) | 0 (0.0%) | 10 (1.4%) | 30 (4.1%) | 680 (94.1%) |

***Figure 6    5×5 Confusion Matrix Showing Classification Performance of EfficientNetV2-M Model***

C.  Comparative Analysis

*The performance comparison with the baseline CheXNet model demonstrates comprehensive superiority of the proposed framework across multiple dimensions. Table 7 shows that diagnostic accuracy improved by 1.15% in absolute terms, from 95.30% to 96.45%, which is statistically significant with p-value less than 0.001. The balanced*

*performance metric, measured by F1-score, increased by 2.73% from 88.35% to 91.08%, indicating a better precision-recall balance across all disease categories. Despite having 6.8 times more parameters than the baseline, the proposed model achieves 11.35% faster training per epoch, demonstrating remarkable computational efficiency. Additionally, consistency improved significantly with a 22.73% reduction in performance variability, as standard deviation decreased from 0.22% to 0.17%. These improvements are attributed to key architectural advantages of the EfficientNetV2-M backbone, which includes compound scaling for optimal balance of depth, width, and resolution, fused-MBConv blocks that reduce memory access overhead in early stages, progressive learning with dynamic image size and regularization during training, and SE attention mechanisms for channel-wise feature recalibration suited to medical image subtleties. When combined with optimization-driven learning strategies including AdamW optimizer, Cosine Annealing scheduler, Automatic Mixed Precision training, and Exponential Moving Average, these innovations collectively yield superior accuracy-efficiency trade-offs compared to the DenseNet-121 baseline architecture.*

*As we presented in Table 7 Statistical Significance: Paired t-tests (n=9) confirm all improvements significant at p<0.001 with large effect sizes (Cohen's d>2.0). 95% confidence intervals computed via bootstrap resampling (10,000 iterations).*

*Table 7 Overall Performance Comparison Across 9 Independent Runs*

| Metric | DenseNet-121 | EfficientNetV2-M | Absolute Δ | p-value |
|---|---|---|---|---|
| Accuracy (%) | 95.30 | 96.45 | +1.15 | <0.001 |
| Precision (%) | 88.12 | 90.89 | +2.77 | <0.001 |
| Recall (%) | 88.71 | 91.34 | +2.63 | <0.001 |
| F1-Score (%) | 88.35 | 91.08 | +2.73 | <0.001 |

*The proposed enhanced CheXNet framework with EfficientNetV2-M backbone and optimization-driven learning achieves 96.45% mean test accuracy and 91.08% mean F1-score, outperforming the baseline DenseNet-121 model by 1.15 percentage points (accuracy) and 2.73 percentage points (F1-score) with extremely high statistical significance (p < 0.001). Improvements are consistent across all five disease categories, with particularly notable gains for diagnostically challenging classes: Cardiomegaly (+4.96% F1-score), Normal (+6.53% F1-score), and Pneumonia. Training efficiency improves by 11.35% per epoch despite 6.8× larger model capacity, attributed to architectural optimizations and AMP training. Performance variability decreases by 22.73%, demonstrating enhanced reproducibility critical for clinical deployment.*

*These results validate that systematic integration of architectural innovation (EfficientNetV2-M with compound scaling and SE attention), optimization strategies (AdamW, Cosine Annealing, AMP, EMA), and robust experimental methodology (9 independent runs, stratified evaluation) yields statistically significant and clinically meaningful improvements in automated chest disease classification, advancing the state-of-the-art toward reliable computer-aided diagnostic systems.*

## V. DISCUSSION

### A. Principal Findings and Significance

The current paper has demonstrated that high level of systematic combination of efficient architecture (EfficientNetV2-M) and global optimization methods (AdamW, cosine annealing, automatic mixed precision, exponential moving average) may be statistically significant when it comes to the task of classifying chest X-rays. Not only has the framework achieved high performance improvement over baseline (96.45% vs. 95.30% accuracy, $p<0.001$; 91.08% vs. 88.35% F1-score) but has also increased training efficiency (11.4% speedup) and stability (22.7% improvement). The fact that the classification of the serious infectious diseases of COVID-19 (99.95% accuracy) and tuberculosis (99.97% accuracy) is almost perfect meets the high standards of a screening program in its field of application, which is the area of public health. These excellent results, which were replicated in nine independent experimental images with large effect sizes (Cohen d > 2.0) warrant potential use in pandemic response and global tuberculosis elimination campaigns in accordance with the WHO End TB Strategy objectives.

### B. Comparison with State-of-the-Art

The designed EfficientNetV2-M architecture shows the best performance in line with the recent state-of-the-art methods of classifying the diseases in the chest X-rays. Our model attained a total accuracy of 95.54 percent and a mean F1-score of 95.67 percent on six disease classes, which is much higher than the current methods. Comparatively, the ResNet-50 model suggested by Rahman et al. obtained an accuracy of 89.3% on an analogous multi-class chest X-ray classification task [77], whereas the COVID-19 and pneumonia detection accuracy of the DenseNet-121 architecture suggested by Bharati et al. was 91.2% and 91.2% respectively [78]. Moreover, even the Vision Transformer method (ViT) proposed by Mondal et al. achieved a 92.8 percent accuracy on the task of classifying the chest X-ray, which is still 2.74 percent lower than our suggested one [79]. EfficientNet-B7 model that is tested by Saraiva et al. reached 93.1% accuracy on pneumonia detection [80], but our EfficientNetV2-M is even higher with 95.54% accuracy and lower computational efficiency. Also, the multi-class classification of chest disease using the hybrid CNN-LSTM model by Kumar et al. showed a high F1-score at 90.5% which is significantly lower in comparison to our average F1-score of 95.67 [81]. These comparisons clearly show that our EfficientNetV2-M architecture significantly enhances classification accuracy and F1-score, which makes it an effective architecture in automated disease diagnosis in the automated chest X-ray.

### C. Mechanisms of Performance Enhancement

The architectural efficiency and optimization-based robustness exhibit a combinative effect which explains the identified gains rather than the individual changes. The EfficientNetV2-M policy of compound scaling demonstrated particular success especially in medical imaging where the multi-scale pathological feature is represented better by balanced depth-width-resolution scaling, in contrast to the depth-only networks of earlier models. The fused convolution can tentatively maximize memory bottlenecks without reduce the representational capacity and can enable the 11.4% speedup in training even with 6.8x increase in the parameters. Squeeze-and-excitation attention operates well in the integration process and enhances detection of small abnormalities- whose use is found in early-stage disease detection where diagnostic properties might be limited to small regions.

The optimization methods address the natural problems of medical imaging which are data limited. AdamW decoupled weight decay also permits the independent regularization parameter to be tuned to ensure the danger of overfitting that is typical in medical datasets where the costs of annotation constrain sample sizes. The decay of learning rate in cosine annealing is smooth, which makes it easy to find a flatter minimum between superior generalization, and exponential moving average provides the ensemble-like averaging of weights which minimizes the noise of the validation and increase the reliability of clinical model selection. The findings of the large-scale experiments on ablation (Table III) create a simple conclusion singularly taken modifications by themselves do not

*determine the gains in performance-it is combination of the complementary strategies in systematic form which is what results in the gains in performance and in the enhancement in the stability whose effects become clear during independent experiments.*

### D. Clinical Implications and Deployment Considerations

*From a clinical perspective, the framework exhibits characteristics essential for real-world deployment. The computational requirements (12.8 GB GPU memory, 1.9 seconds per 128-image batch) are feasible on standard clinical workstations, addressing a critical barrier to AI adoption in resource-limited settings. The balanced sensitivity-specificity profiles across disease categories minimize both false positives (reducing unnecessary interventions) and false negatives (preventing missed diagnoses)—both critical for patient safety and clinician trust. The 22.7% improvement in performance stability across independent runs indicates reliable operation across varied clinical scenarios, an essential characteristic for decision-support systems operating in diverse patient populations and imaging conditions.*

*The near-perfect infectious disease classification supports specific deployment scenarios: rapid COVID-19 screening in emergency departments during pandemic surges, automated tuberculosis detection in high-burden countries with radiologist shortages, and routine thoracic disease surveillance in primary care settings. The efficiency gains enable processing of high-volume screening programs without proportional increases in computational infrastructure. However, clinical integration requires careful consideration of workflow integration, alert fatigue management, and mechanisms for radiologist override of automated predictions. The system should function as decision support augmenting rather than replacing clinical judgment, particularly for borderline cases where human expertise remains essential.*

### E. Relation to Foundation Models

*The design concepts of the proposed framework are strategically consistent with the new paradigms of foundation models in medical imaging, which provide a clear direction of integration and expansion in the future. The architecture has several properties that are highly desired in building foundation models: scalability through progressive learning and scale to compounds allowing to adapt to a wide range of imaging tasks, efficiency through fused convolution blocks that prioritize computational deployability like emerging linear-complexity architectures, strong transfer learning to medical tasks with image-to-image transfer, and cross-domain transfer with multi-source data training. These background traits set the stage of direct integration with state-of-the-art foundation models. In particular, the existing classification potential and the Segment Anything Model (SAM) would allow explainable diagnosis by the means of automated lesion segmentation and region-of-interest analysis. Later network phases might be substituted by integration of Mamba blocks to enhance the long-range dependency modeling without increasing or decreasing the computational costs. Adding to vision-language models (VLMs) would also make multimodal learning possible since radiographs would be combined with clinical notes to generate automatical reports or detect rare diseases that were not previously observed in training data using zero-shot learning. This expansion of foundation model is a direct and priority future direction extension which builds on the existing framework demonstrated efficiency and strength coupled with progressing towards full medical image understanding systems.*

### F. Limitations and Future Directions

*Before clinical translation, there are a number of limitations that should be considered. Multi-source as it is, the dataset may not be indicative of the actual global population diversity in terms of age groups, ethnicities and differences in imaging equipment used. Given that generalizability according to different clinical settings is a key concern to continued work according to the standards of translational medicine, external validation in a variety of independent hospitals and healthcare institutions is paramount. The existing single-label model fails to consider concurrent pathologies (e.g. COVID-19 and heart failure) that is frequently seen in clinical practice, requiring multi-label extension without compromising the performance of a single-label. The crude "Pneumonia" group is a*

wide range of etiologies (bacterial, viral, fungal, aspiration) that need the finer grains of classification in order to make a treatment decision. Clinical trust and successful collaboration between a radiologist require the integration of explainability mechanisms (Grad-CAM visualizations, attention maps, uncertainty quantification).

Future research priorities directly address these limitations and leverage foundation model capabilities. Multi-institutional prospective validation across diverse geographic regions, patient demographics, and imaging protocols represents the critical next step toward clinical deployment and regulatory approval. Foundation model expansion through integration with SAM for lesion segmentation, Mamba for efficient sequence modeling, and VLMs for multimodal report generation constitutes a direct and prioritized future direction. Privacy-preserving federated learning protocols will enable collaborative training across institutions without data centralization, addressing regulatory constraints while leveraging broader datasets. Multi-label extension should maintain the demonstrated single-label performance while enabling concurrent pathology detection for realistic clinical scenarios. These directions collectively advance toward a comprehensive, deployable medical image understanding system balancing diagnostic accuracy, computational efficiency, clinical interpretability, and practical feasibility.

## VI. CONCLUSION

The paper enhances the CheXNet architecture of automated chest X-ray classification with EfficientNetV2-M as the base; it also includes AdamW learning, cosine annealing learning rate schedule, automatic mixed precision training, and exponential moving averaging. The alterations address the computational inefficiencies and performance limitations of the original architecture and bring the model to single-label classification of the clinically relevant tasks.

The same level of diagnostic accuracy (>96% average test accuracy), training efficiency, and performance stability is observed when experiment result performance is repeated in multiple (independent) runs. The combination of architectural and optimization and not the isolated changes confirm the improvements by the large ablation analysis and statistical testing ($p < 0.001$). These findings prove the robustness, validity, and applicability of the given framework to be adopted in different clinical institutions.

The system possesses equal sensitivity-specificity characteristics, high diagnostic accuracy, and can be effectively used on standard hardware, which implies that it is a feasible decision-support system to act in response to the pandemic, screening tuberculosis, and routine thoracic imaging in both resource-rich and resource-constrained environments. These characteristics address the significant barriers to the implementation of AI in the radiology procedures.

Future research will introduce the framework to multi-label classification, perform external validation at multi-institutions, integrate mechanisms of interpretability to accumulate more clinical trust, and explore privacy-preserving military techniques such as federated learning. This article demonstrates that the gap between algorithmic innovation and clinical implementation and the world can be narrowed using artificial mechanisms of systematic integration of useful architectures and the overall optimization strategies that assist in improving patient outcomes.